\documentclass[twocolumn,float]{elsart}
\usepackage{graphicx}
\usepackage{amssymb,amsmath}
\usepackage{pslatex}

\begin{document}
\journal{Journal of non-crystalline solids}

\begin{frontmatter}
\title{Asymptotic Description of Schematic Models for CKN}
\author{Matthias Sperl}
\address{Duke University, Department of Physics, Box 90305,
Durham, NC 27708, USA}

\begin{abstract}

The fits of $0.4 Ca(NO_3)_2 0.6 K(NO_3)$ (CKN) by schematic mode-coupling 
models [V. Krakoviack and C. Alba-Simionesco, J. Chem. Phys. \textbf{117}, 
2161-2171 (2002)] are analyzed by asymptotic expansions. The validity of 
both the power-law and the Cole-Cole-peak solutions for the critical 
spectrum are investigated. It is found that the critical spectrum derived 
from the fits is described by both expansions equally well when both 
expansions are carried out up to next-to-leading order. The expansions up 
to this order describe the data for 373K over two orders of magnitude in 
frequency. In this regime an effective power law $\omega^a$ can be 
identified where the observed exponent $a$ is smaller than its calculated 
value by about 15\%; this finding can be explained by corrections to the 
leading-order terms in the asymptotic expansions. For higher temperatures, 
even smaller effective exponents are caused by a crossover to the 
alpha-peak spectrum.

\end{abstract}

\end{frontmatter}

\section{Introduction}

For a given interaction potential mode-coupling theory (MCT) in its 
microscopic version can predict the transition from a fluid to a glass 
\cite{Bengtzelius1984}. Close to the transition, the glassy dynamics is 
ruled by universal scaling laws, that are independent of the details of 
the interactions. These universal laws have been applied frequently to 
describe experimental data, especially when the microscopic details are 
difficult to model, which is the case in most molecular glass formers. At 
larger distances from the liquid-glass transition, non-universal 
corrections to scaling become important \cite{Franosch1997}. These can be 
calculated exactly by asymptotic expansion for a given MCT model, but 
introduce additional parameters for the fitting of data. An alternative to 
fitting higher-order terms of an asymptotic expansion is provided by 
schematic MCT models. These simplified models incorporate naturally both 
universal features and corrections, and can be used to fit data by a 
limited number of parameters. Similar to the microscopic models, 
asymptotic expansions can also be devised for the schematic models for 
experimentally relevant parameters, which in turn provides insight into 
the validity of the universal laws and the importance of the corrections 
for the data under investigation.

One particular schematic model used frequently is the $F_{12}$ model which 
deals with a single correlator mimicking the density-correlators 
\cite{Goetze1984}. This model can be supplemented by a second correlator 
to describe some specific probe variable \cite{Sjoegren1986}; and this 
two-correlator model has been applied successfully to the description of 
experimental data 
\cite{Singh1998,Ruffle1999,Goetze2000b,Brodin2002,Wiebel2002,Cang2005}. 
Data for salol \cite{Hinze2000} and benzophenone (BZP) \cite{Cang2003} 
were fitted recently by this model, and it was subsequently shown that an 
apparent violation of universal MCT formulas \cite{Ricci2002,Cang2003} can 
be reconciled with a more detailed asymptotic analysis within MCT 
\cite{Goetze2004,Goetze2005}: A Cole-Cole peak found earlier for this 
schematic model \cite{Buchalla1988,Goetze1989c} was identified in the fits 
for these two substances. While the asymptotic approximation by the 
Cole-Cole peak improved the description of salol only quantitatively, the 
presence of a Cole-Cole peak was essential for interpreting the BZP 
spectra even qualitatively. A remarkable wing in the Giga-Hertz regime 
could be explained by a significant contribution of the Cole-Cole peak 
\cite{Goetze2005,Cummins2005}.

Another substance studied in detail is the molten salt $0.4 Ca(NO_3)_2 0.6 
K(NO_3)$ (CKN) which was measured in light-scattering experiments 
\cite{Li1992}. These data were successfully fitted by a number of 
different schematic models \cite{Krakoviack1997,Krakoviack2002}. For these 
models the asymptotic expansions will be derived in an effort to identify 
asymptotic features present in the data of CKN. The models will be 
introduced in Sec.~\ref{sec:models}, their asymptotic solutions presented 
in Sec.~\ref{sec:asy}; Sec.~\ref{sec:results} discusses the two lowest 
temperatures considered in \cite{Krakoviack2002}, 
Sec.~\ref{sec:discussion} extends the analysis down to temperatures as low 
as in Ref.~\cite{Krakoviack1997}, and presents a conclusion.

\section{Schematic Model Fits\label{sec:models}}

\begin{figure}
\includegraphics[width=\columnwidth]{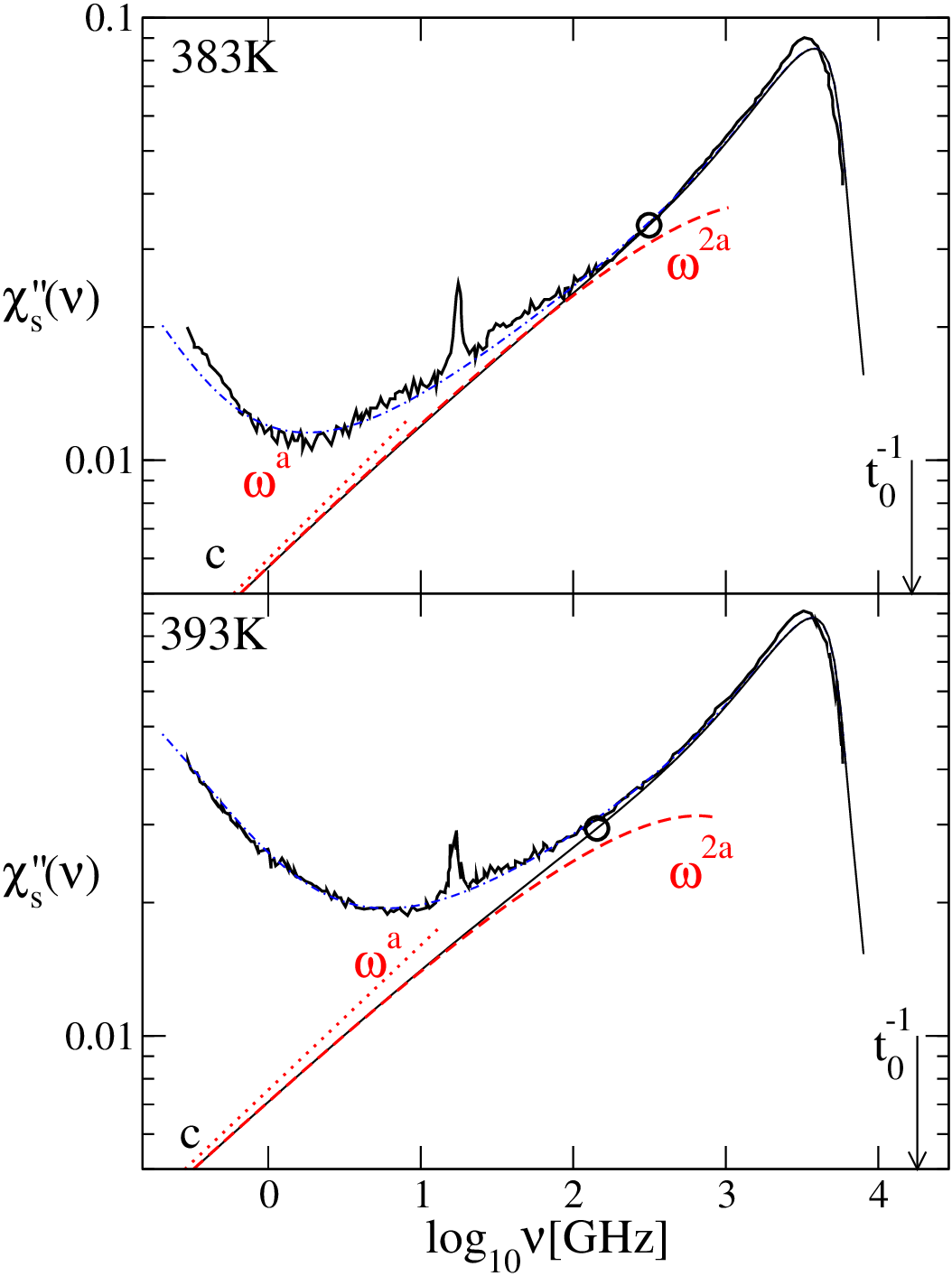}
\caption{\label{fig:ckn}Power-law expansion for CKN. The full curves 
reproduce the data from Ref.~\cite{Li1992}. The dashed-dotted curves 
present schematic-model fits using kernel~(\ref{eq:mr}) and the effective 
correlator~(\ref{eq:scatter_22}) with $r=8.6$ as in \cite{Krakoviack2002}. 
The full lines labeled \textit{c} display the critical spectra for 
$\lambda = 0.656$ ($a = 0.344$, upper panel) and $\lambda = 0.7$ ($a = 
0.327$, lower panel), respectively. The dotted lines labeled $\omega^a$ 
and the dashed lines labeled $\omega^{2a}$ show the leading and 
next-to-leading order result of Eq.~(\ref{eq:chi_power:chi}). The points 
where the critical spectra deviate by 10\% from the approximation 
by (\ref{eq:chi_power:chi}) are indicated by circles.
}
\end{figure}

MCT provides equations of motion for correlation functions $\phi_q(t)$ 
where the index $q$ denotes the wave-number or a specific dynamical 
variable. With the initial conditions $\phi_q(t=0) = 1$ and 
$\partial_t\phi_q(t=0)=0$, the MCT equations read
\begin{equation}\label{eq:EOM:int}
\begin{split}
\partial_t^2 \phi_q(t) + \nu_q\partial_t\phi_q(t)
+  \Omega_q^2 \phi_q(t)\hfill \\
+ \Omega_q^2 \int_0^t\,dt'\,m_q(t-t')\partial_{t'}\phi_q(t')
= 0\,,
\end{split}
\end{equation}
where $\Omega_q$ and $\nu_q$ specify normal liquid dynamics. In the 
$F_{12}$ model, the dynamics of the density correlators is modeled by a 
single correlation function with kernel
\begin{equation}\label{eq:mF12}
m(t) = v_1\phi(t) + v_2 \phi(t)^2\,.
\end{equation}
The probing variable is given by a second correlator $\phi_A(t)$ and a
corresponding kernel \cite{Sjoegren1986},
\begin{equation}\label{eq:mA}
m_A(t) = v_A \phi(t) \phi_A(t)\,.
\end{equation}
The dynamics of $\phi_q(t)$  is given by Eq.~(\ref{eq:EOM:int}) with $q$ 
replaced by $A$. A different kernel for a second correlator was introduced 
by Alba-Simionesco and coworkers that shall be denoted by $\phi_r(t)$ 
\cite{Alba1995},
\begin{equation}\label{eq:mr}
m_r(t) = r m(t) = r\left[ v_1\phi(t) + v_2 \phi(t)^2\right]\,.
\end{equation}
In addition, a number of effective correlators $\phi_s(t)$ were defined, 
some of whose were originally motivated to capture the DID mechanism 
\cite{Alba1995,Krakoviack2002},
\begin{subequations}\label{eq:scatter}
\begin{eqnarray}
\label{eq:scatter_22}
\phi_s(t) 	&=&	\gamma \phi(t)^2 + (1-\gamma) \phi_{A,r}(t)^2\,,\\
\label{eq:scatter_21}
\phi_s(t) 	&=&	\gamma \phi(t)^2 + (1-\gamma) \phi_{A,r}(t)\,,\\
\label{eq:scatter_12}
\phi_s(t) 	&=&	\gamma \phi(t)   + (1-\gamma) \phi_{A,r}(t)^2\,,\\
\label{eq:scatter_11}
\phi_s(t) 	&=&	\gamma \phi(t)   + (1-\gamma) \phi_{A,r}(t)\,.
\end{eqnarray}
\end{subequations}
Here, $\phi_{A,r}(t)$ denotes a second correlator with the kernel from 
Eq.~(\ref{eq:mA}) or Eq.~(\ref{eq:mr}), respectively. The data was fitted 
with the susceptibility given by the Fourier transformation
\begin{equation}\label{eq:chi}
\chi_s''(\omega) = \omega\int_0^\infty\,dt\,\cos(\omega t) \phi_s(t)\,.
\end{equation}
Further details and the values of the fit parameters can be found in 
Refs.~\cite{Krakoviack2002,Krakoviack2000a}. The fit curves are reproduced 
in Fig.~\ref{fig:ckn} together with the experimental data for CKN for the 
two lowest temperatures considered in \cite{Krakoviack2002}. Here,
kernel~(\ref{eq:mr}) and the effective correlator~(\ref{eq:scatter_22}) 
are applied; the frequencies are denoted in units of $\nu = 
\omega/(2\pi)$.

\section{Asymptotic Solution for Schematic Models\label{sec:asy}}

The general asymptotic formulas for the critical spectrum are given in the 
following and are specialized to the schematic models subsequently. From 
those formulas the asymptotic description of CKN shall be inferred. The 
following discussion is restricted to spectra directly at the transition 
point, which will be called critical spectra. Such a critical spectrum is 
shared by larger and larger parts of the dynamics when the temperature is 
approaching the critical temperature $T_c$. Two different asymptotic 
approximations for the critical spectrum shall be used, a well-known 
power-law expansion and a modified Cole-Cole law introduced recently.

The power-law solution for the critical spectrum is derived from the 
asymptotic expansion for the correlators. It reads up to next-to-leading 
order \cite{Franosch1997},
\begin{subequations}\label{eq:chi_power}
\begin{equation}\label{eq:chi_power:chi}\begin{split}
\chi_q''(\omega) = \left[ \Gamma(1-a) \sin\left( \pi a/2 \right) \right] 
h_q (\omega t_0)^a \\\times
\left[ 1+ k_a \hat{K}_q (\omega t_0)^a \right]\,,
\end{split}\end{equation}
with the prefactor
\begin{equation}\label{eq:chi_power:ka}
k_a =  2 \Gamma(1-a) \cos\left( \pi a/2 \right)/\lambda\,,
\end{equation}
the critical amplitude $h_q$, the critical exponent $a$ -- given by 
the exponent parameter $\lambda$ using the Euler-$\gamma$ function in
$\lambda = \Gamma(1-a)^2/\Gamma(1-2a)$ -- the microscopic time scale 
$t_0$, and a correction amplitude $\hat{K}_q$. The first line in 
Eq.~(\ref{eq:chi_power:chi}) constitutes the leading-order result: an 
increase of the critical spectrum with as $\omega^a$.
\end{subequations}

In contrast to Eq.~(\ref{eq:chi_power}), the modified Cole-Cole law is 
obtained from using the expansion for the kernel $m_q(t)$ rather than the 
one for $\phi_q(t)$ \cite{Goetze2005}:
\begin{subequations}\label{eq:cccpar}
\begin{equation}\label{eq:ccc}\begin{split}
\chi_q(\omega) = \chi_{0\,q}^{cc} /\left[ 1 + \left( 
-i\omega/\omega_q^{c}  \right)^a \right.\\\left.
+ \hat{K}^{cc}_q  \left( 
-i\omega/\omega_q^{c}  \right)^{2a}\right] \,,
\end{split}\end{equation}
where all parameters can be cast in a form using only results known from 
the power-law expansion. The amplitude is given by 
\begin{equation}\label{eq:cccpar:chi0}
\chi_{0\,q}^{cc} = 1- f^c_q\,,
\end{equation}
with the glass-form factor $f_q^c$; the characteristic frequency reads
\begin{equation}\label{eq:cccpar:wcc}
\omega_q^{c}t_0 = \left[ (1-f^c_q) / \left(h_q \Gamma(1-a)\right)  
\right]^{1/a}\,,
\end{equation}
and the correction amplitude is expressed as 
\begin{equation}\label{eq:cccpar:cqcc}
\hat{K}_q^{cc} = 1 + \left[(1-f^c_q)/h_q\right] \hat{K}_q / \lambda\,. 
\end{equation}
For $\hat{K}_q^{cc} = 0$, the correction vanishes, and the leading-order 
result in Eq.~(\ref{eq:ccc}) is recovered as the original Cole-Cole 
function \cite{Cole1941}.
\end{subequations}

For the schematic models used in the following, one can rewrite all
the parameters needed in Eqs.~(\ref{eq:chi_power}) and (\ref{eq:cccpar}) 
in terms of only two variables: the exponent parameter $\lambda$ and the 
respective coupling parameters, $v_A$ or $r$. We obtain for the glass-form 
factors,
\begin{subequations}\label{eq:pars}
\begin{equation}\label{eq:pars_f}\begin{split}
f^c = 1-\lambda\,,\quad
f_A =& \frac{v_A(1-\lambda)-1}{v_A(1-\lambda)}\,,\\
f_r =& \frac{r(1-\lambda)}{r(1-\lambda)+\lambda}\,,
\end{split}\end{equation}
for the critical amplitudes,
\begin{equation}\label{eq:pars_h}\begin{split}
h   = \lambda\,,\quad
h_A =& \lambda/\left[v_A (1-\lambda)^2\right]\,,\\
h_r =& \frac{r\lambda}{\left[r(1-\lambda)+\lambda\right]^2}\,,
\end{split}\end{equation}
and for the correction amplitudes,
\begin{equation}\label{eq:pars_K}\begin{split}
\hat{K}  =& \kappa\,,\\
\hat{K}_A=& \kappa + \frac{\lambda}{1-\lambda}\left[
\frac{v_A(1-\lambda)}{v_A-1/(1-\lambda)}-1\right]
\,,\\\
\hat{K}_r=& \kappa + \lambda^2\frac{1-r}{r(1-\lambda)+\lambda}\,.
\end{split}\end{equation}
$\kappa$ can be determined straightforwardly from the exponent parameter 
$\lambda$ \cite{Goetze1989c}. For the discussion below, $\kappa$ will 
always contribute only a very small part to the correction amplitude; 
assuming $\kappa = 0$ would not alter any results significantly.
\end{subequations}

The models for the second correlators in Eqs.~(\ref{eq:mA}) and 
(\ref{eq:mr}) are in fact very similar, and the preceeding formulas also 
motivate a closer comparison. From setting $f_A=f_r$, one obtains a 
mapping between the coupling coefficients,
\begin{equation}\label{eq:alba_mapping}
r = v_A\lambda-\frac{\lambda}{1-\lambda}\,,\quad
v_A = \frac{r}{\lambda}+\frac{1}{1-\lambda}\,.
\end{equation}
When $v_A$ and $r$ are related by Eq.~(\ref{eq:alba_mapping}), the ratio 
between the critical amplitudes becomes
\begin{equation}\label{eq:alba_mapping_h}
\frac{h_A}{h_r} = \lambda + \frac{\lambda^2}{r(1-\lambda)}\,,
\end{equation}
which becomes unity for $r = \lambda^2/(1-\lambda)^2$. For that specific 
value of $r$, the mapping~(\ref{eq:alba_mapping}) is correct in leading 
asymptotic order: $\chi_A''(\omega)$ and $\chi_r''(\omega)$ have the same 
leading-order approximation in both Eq.~(\ref{eq:chi_power:chi}) 
and (\ref{eq:ccc}). Imposing again Eq.~(\ref{eq:alba_mapping}), the ratio 
among the correction amplitudes is 
\begin{equation}\label{eq:alba_mapping_K}
\frac{\hat{K}_A-\kappa}{\hat{K}_r-\kappa} = 1 + 
\frac{\lambda}{r(1-\lambda)}\,,
\end{equation}
which cannot be unity with $h_a/h_r$ simultaneously. However, the 
correction amplitudes get closer to each other for larger values of $r$. 
Typical values for $r$ and $v_A$ can be obtained from 
Ref.~\cite{Krakoviack2002}; there, both models for the second correlator 
were used for fitting the CKN data independently. The values reported in 
Ref.~\cite{Krakoviack2002} agree within a 15\% margin with the mapping in 
Eq.~(\ref{eq:alba_mapping}); the values for $r$ also allow for an 
approximate identification of the amplitudes, $h_A\approx h_r$, and the 
ratio in Eq.~(\ref{eq:alba_mapping_K}) lies between a maximum of 2 and a 
minimum as low as 1.2. Two conclusions can be drawn from the preceeding 
discussion. First, both models, $\chi_A''(\omega)$ and $\chi_r''(\omega)$, 
are almost equivalent in their asymptotic properties for experimentally 
relevant parameter values; and second, the fitting procedures employed in 
Ref.~\cite{Krakoviack2002} show this equivalence consistently.

The asymptotic expansions for the effective correlators and their spectra
are calculated by inserting the results for $\phi(t)$ and $\phi_{A,r}$ into 
Eq.~(\ref{eq:scatter}) and retaining only terms up to second order. For 
the case (\ref{eq:scatter_22}) one gets, 
\begin{equation}\label{eq:asy_s}\begin{split}
f_s = \gamma f^{c\,2} + (1-\gamma)f_{A,r}^2\,,\\
h_s = 2\gamma f^c h + 2(1-\gamma)f_{A,r}h_{A,r}\,,\\
\hat{K}_s = \left[\gamma (h^2+2 f^c h \hat{K}) 
	+ (1-\gamma) \right.\\\left.\times(h_{A,r}^2+ 2 f_{A,r} h_{A,r} 
	\hat{K}_{A,r})\right]/h_s\,.
\end{split}
\end{equation}
The remaining cases from Eq.~(\ref{eq:scatter}) can be calculated in the 
same way and yield similar formulas, so these results need not be given 
here. The parameters for the Cole-Cole law result from inserting the 
values from Eq.~(\ref{eq:asy_s}) into Eq.~(\ref{eq:cccpar}) with $q = s$.

\section{Results\label{sec:results}}

The results shown in this section will be restricted to temperatures at 
and above 383K as in \cite{Krakoviack2002}. There, this restriction was 
motivated to avoid the artefacts in the original data \cite{Li1992} 
identified for lower temperatures 
\cite{Surovtsev1998,Gapinski1999,Barshilia1999}. However, the results 
derived for the critical spectrum in the present section will be shown to 
be relevant also for lower temperatures in Sec.~\ref{sec:discussion}. For 
the discussion of the critical spectrum, an appropriate point on the 
transition surface of the $F_{12}$ model needs to fixed. In the $F_{12}$ 
model such a critical point can be specified uniquely by the value of 
$\lambda$. One reasonable choice is a transition point close to the fit 
parameters for the lowest temperature under discussion. Such a point is 
chosen for the critical spectrum shown in the upper panel of 
Fig.~\ref{fig:ckn}; the exponent parameter for this critical point is 
$\lambda = 0.656$. Another way to determine a suitable transition point is 
suggested by an extrapolation procedure in Ref.~\cite{Krakoviack2002}, 
where $\lambda \approx 0.7$ is proposed for the description of the 
dynamics. The critical spectrum for the transition point satisfying 
$\lambda = 0.7$ is added to the lower panel of Fig.~\ref{fig:ckn}; to show 
also experimental results close to that transition point, data and fit of 
the nearby 393K spectrum is added.

The asymptotic approximation by Eq.~(\ref{eq:chi_power:chi}) is shown in 
Fig.~\ref{fig:ckn}, and it is seen clearly that the leading-order power 
law describes the critical spectrum only for $\nu\lesssim 1$GHz, 
regardless of the chosen transition point. Including the correction 
describes the critical solutions up to 300GHz and 150GHz, respectively. 
However, Eq.~(\ref{eq:chi_power:chi}) explains the measured spectrum for 
383K for only about one order of magnitude in frequency. The apparent 
power-law behavior $\nu^{a'}$ with $a' = 0.22$ between 2GHz and 200GHz for 
383K does in turn not show the asymptotic value $a=0.344$ for the critical 
exponent. This is because the separation of this state point from the 
transition is too large.

\begin{figure}
\includegraphics[width=\columnwidth]{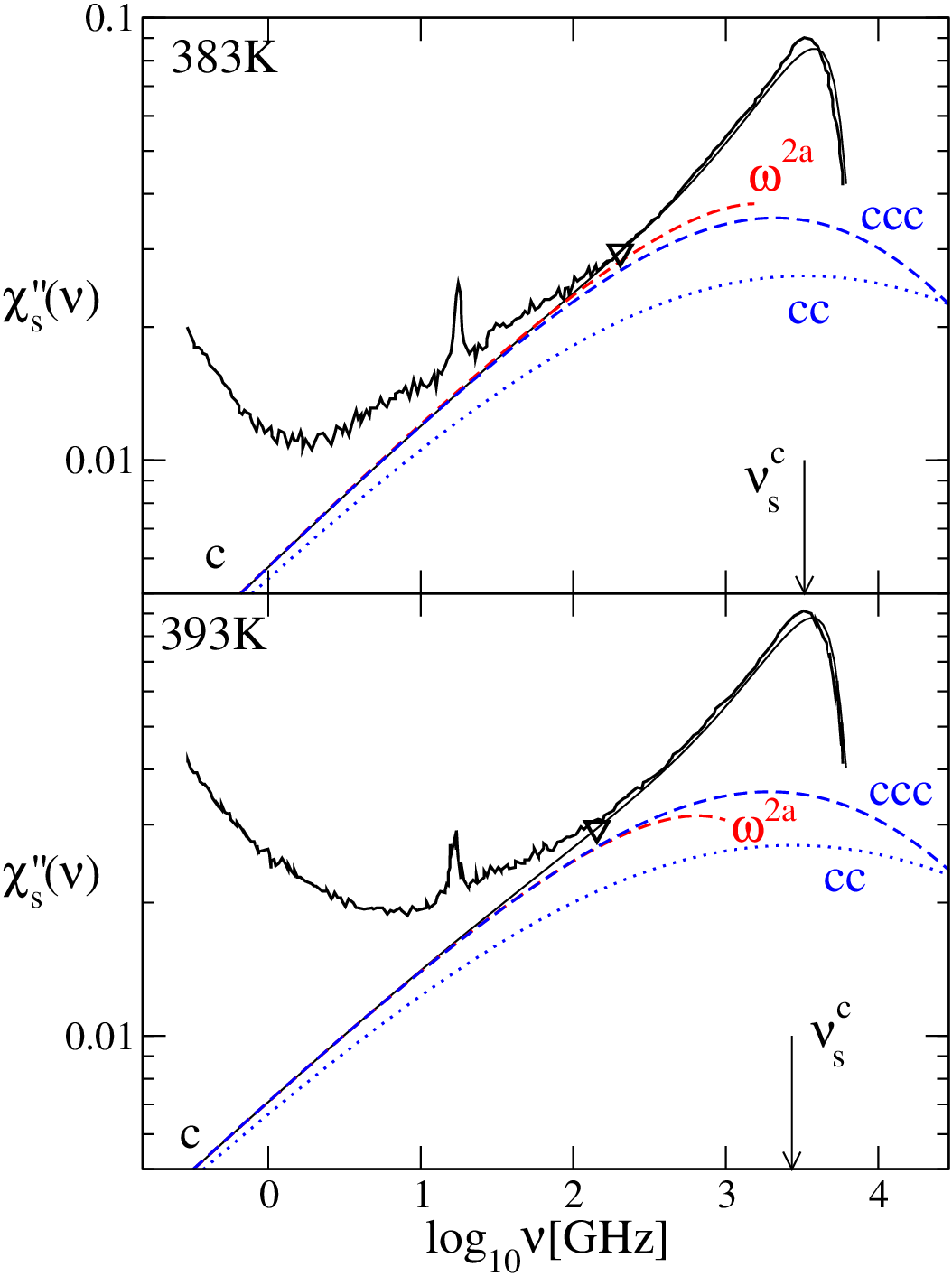}
\caption{\label{fig:ckncc}Cole-Cole law for CKN. The full curves display 
the experimental data \cite{Li1992} as before. Also reproduced from 
Fig.~\ref{fig:ckn} are the respective critical spectra (\textit{c})
and the power-law approximations by Eq.~(\ref{eq:chi_power}) up to 
next-to-leading order ($\omega^{2a}$). The dotted line labeled 
\textit{cc} and the dashed line labeled \textit{ccc} show the leading- and 
next-to-leading order approximation by the Cole-Cole law, 
Eq.~(\ref{eq:ccc}) for frequencies indicated by the arrows. A 10\% 
deviation of the critical spectra from their respective approximation by 
Eq.~(\ref{eq:ccc}) are shown by the triangles.
}
\end{figure}

The alternative asymptotic approximation by the Cole-Cole law, 
Eq.~(\ref{eq:ccc}), is shown in Figure~\ref{fig:ckncc}. Similar to the 
power-law solution, the leading-order Cole-Cole formula applies to the 
critical spectra only for $\nu < 1$GHz, but has no relevance for the 
description of the present data. The formula including the correction 
describes the critical spectra in upper and lower panel for $\nu < 200$GHz 
and $\nu < 150$GHz, respectively. The comparison of both panels also 
reveals that the choice of a particular transition point does not lead to 
major differences. In both cases, Cole-Cole and power-law expansion are 
practicably indistinguishable for $\nu\lesssim 500$GHz. This is also the 
regime where they both approximate the critical spectrum well.

\begin{table*}[ht]
\begin{tabular}{lllllll}
	& I & II & III &IV & V &VI\\\hline
$\phi_s$& 
(\ref{eq:mr}), (\ref{eq:scatter_22}) &
(\ref{eq:mr}), (\ref{eq:scatter_21}) &
(\ref{eq:mr}), (\ref{eq:scatter_12}) &
(\ref{eq:mr}), (\ref{eq:scatter_11}) &
(\ref{eq:mA}), (\ref{eq:scatter_22}) & 
(\ref{eq:mA}), (\ref{eq:scatter_11})\\
$\lambda$ & 0.656 & 0.644& 0.642& 0.629& 0.630& 0.644\\
$\hat{K}_s$ & -0.31& -0.24&-0.32 & -0.28& -0.26& -0.32\\
$\nu^c_s$[GHz] &3280 &3060 &2650 &2980 &2750 & 2410\\
$\chi^{cc}_{0\,s}$ & 0.496& 0.491&0.483 &0.478 &0.510 & 0.507\\
$\hat{K}_s^{cc}$ & 0.63& 0.68& 0.67& 0.68&0.67 &0.59 \\
\hline
\end{tabular}\caption{\label{tab:IVI}Selected parameter values for the 
asymptotic approximation of the critical spectra. The designation of the 
cases I--VI is the same as in \cite{Krakoviack2002}; the first line shows 
the definition of the effective correlator $\phi_s$.}
\end{table*}

As seen in Fig.~\ref{fig:ckn}, the range of validity for approximation 
(\ref{eq:chi_power:chi}) increases by two decades when the corrections are 
included, $\hat{K}_s = -0.31$ for the upper panel; such a value for 
$\hat{K}_s$ can be considered moderate compared to BZP and salol where the 
corresponding values are around five times larger \cite{Goetze2005}. For 
the lower panel $\hat{K}_s = -0.44$. On the contrary in 
Fig.~\ref{fig:ckncc}, the corrections in (\ref{eq:cccpar:cqcc}) for 
approximation (\ref{eq:ccc}) are much larger for CKN, $\hat{K}^{cc}_s = 
0.63$ and $\hat{K}^{cc}_s = 0.54$ for upper and lower panel, than for BZP 
and salol where the corrections are always close to zero for the Cole-Cole 
law. Table~\ref{tab:IVI} lists the values of characteristic parameters for 
the models used in Ref.~\cite{Krakoviack2002}. Model~I is the one 
discussed in detail above for Figs.~\ref{fig:ckn} and \ref{fig:ckncc}. 
Most parameters are remarkably similar throughout all the models. The 
location of the Cole-Cole peak $\nu_s^c$ varies by only 30\%, so the 
scenario shown in Fig.~\ref{fig:ckncc} is typical for all these models. 
The values for the case $\lambda=0.7$, corresponding to the lower panels 
in Figs.~\ref{fig:ckn} and~\ref{fig:ckncc}, are very similar and are 
therefore omitted.

\section{Discussion\label{sec:discussion}}

\begin{figure}
\includegraphics[width=\columnwidth]{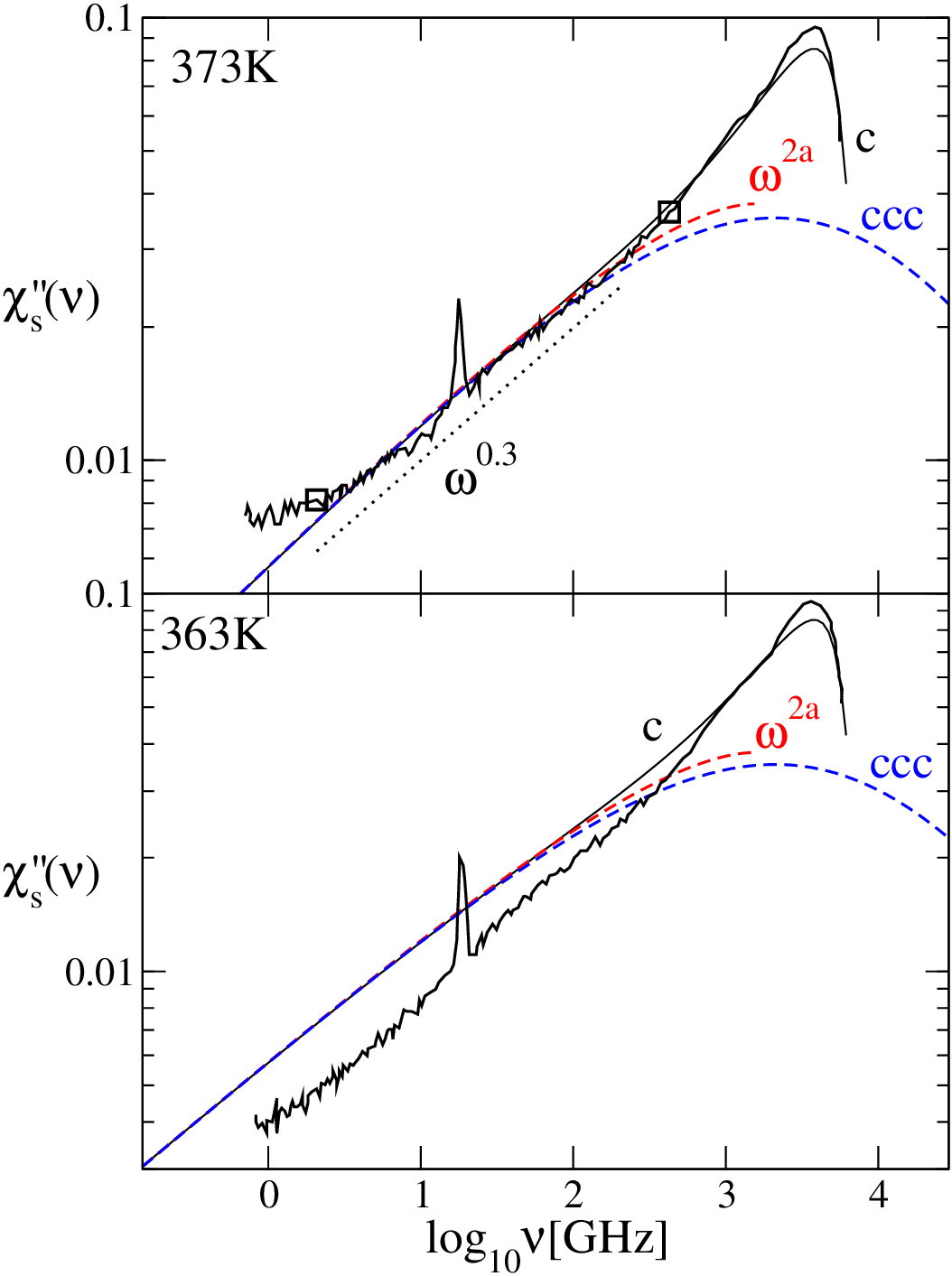}
\caption{\label{fig:ckncrit}Spectra for 373K and 363K. Critical spectrum 
(\textit{c}), power-law approximation ($\omega^{2a}$) and Cole-Cole law 
(\textit{ccc}) are reproduced from Figs.~\ref{fig:ckn} and 
\ref{fig:ckncc}. In the upper panel, the squares indicate a 10\% deviation 
of the approximation~(\ref{eq:chi_power}) from the data. An effective 
power law $\omega^{0.3}$ is shown as dotted line.
}
\end{figure}

Figures~\ref{fig:ckn} and \ref{fig:ckncc} indicate that both the 
approximations, Eq.~(\ref{eq:chi_power:chi}) and (\ref{eq:ccc}), as well 
as the full critical spectra do not describe significant parts of the 
measured data for $T \geqslant 383$K. For lower temperatures, two 
additional spectra with fits from Ref.~\cite{Krakoviack1997} are 
available, $T = 373$K and $T = 363$K. The data for these temperatures from 
Ref.~\cite{Li1992} are shown in Fig.~\ref{fig:ckncrit}. It is apparent 
from the upper panel of Fig.~\ref{fig:ckncrit} that the spectrum for $T = 
373$K coincides with the critical spectrum for 2.5 decades; the asymptotic 
expansions describe the data for over two orders of magnitude. 

The corrections to the power law, cf. Eq.~(\ref{eq:chi_power:chi}), lead 
to a renormalization of the asymptotic power law $\omega^{0.344}$ to an 
effective power law $\omega^{0.3}$. In contrast to the findings for salol 
and BZP, the Cole-Cole formula~(\ref{eq:ccc}) is not essential for 
understanding these smaller exponents for CKN 
\cite{Goetze2004,Goetze2005}. The asymptotic expansions also reveal that 
the minimum seen in the data for CKN can be regarded as a $\beta$-minimum 
with a von~Schweidler wing for lower and a critical spectrum for higher 
frequencies. Therefore, for temperatures $T \gtrsim 383$K, earlier 
analyses of the minimum can be justified \cite{Cummins1993}.

For the 373K spectrum a fit can be found with the same parameters as for 
$T=383$K by adjusting $v_1$ and $v_2$ only. The resulting state is in fact 
close to the critical point for $\lambda = 0.656$, and a similar point was 
found earlier for the same temperature for slightly different parameters 
\cite{Krakoviack1997}. The situation for $T = 363$K is different. The 
spectrum shown in the lower panel of Fig.~\ref{fig:ckncrit} deviates 
drastically from the critical spectrum for $\nu < 1$THz. Hence, in this 
case also the approximations to the critical spectrum are not applicable. 
Although the data for $T = 363$K could be fitted in \cite{Krakoviack1997}, 
this was possible only by moving away from the path for higher 
temperatures rather considerably \cite{Krakoviack2002}. One might 
interpret that as an indication of the mentioned experimental artefacts 
\cite{Krakoviack2002,Surovtsev1998,Gapinski1999,Barshilia1999}. 

For temperatures $T \leqslant 363$K, the exponents reported in Fig.~4 of 
Ref.~\cite{Gapinski1999} are all equal or larger than 0.4. These 
temperatures, when interpreted within MCT, are therefore located below 
$T_c$. This is a consequence of the evolution of the spectra as seen for 
higher temperatures, as well as the asymptotic analysis at the critical 
point: Both mechanisms only allow for \textit{smaller} effective exponents 
compared to the asymptotic value for $a$ which cannot be larger than 
0.396. Below $T_c$, larger effective exponents are not in contradiction to 
MCT; however, they do not describe a critical spectrum. 

In conclusion, for $T \leqslant 363$K, in the spectra from \cite{Li1992} 
as well as \cite{Gapinski1999} no critical MCT spectrum can be 
demonstrated; $T = 383$K marks the highest temperature where part of the 
critical spectrum is observable; and the data for $373$K, cf. upper panel 
of Fig.~\ref{fig:ckncrit}, exhibit the critical MCT spectrum rather 
convincingly.

{\sc Acknowledgements} I want to thank V.~Krakoviack for providing the fit 
parameters from Refs.~\cite{Krakoviack2002,Krakoviack2000a}. Discussions 
with W.~G\"otze and V.~Krakoviack are gratefully appreciated. This work 
was supported by DFG Grant No. SP~714/3-1 and SP~714/5-1, and NSF grants 
DMR0137119 and DMS0244492.

\bibliographystyle{elsart-num}

\end{document}